\documentclass[twocolumn]{aastex62}

\received{xxxx, 2018}
\revised{xxxx, 2018}
\accepted{November 7, 2018}

\shorttitle{Suppression of Low-mass Galaxy Formation around Quasars at $z\sim2-3$}
\shortauthors{Uchiyama et al.}

\usepackage{here}

\begin{document}
\title{Suppression of Low-mass Galaxy Formation around Quasars at $z\sim2-3$}
\correspondingauthor{Hisakazu Uchiyama}
\email{hisakazu.uchiyama@nao.ac.jp}

\author{Hisakazu Uchiyama}
\affiliation{Department of Astronomical Science, SOKENDAI (The Graduate University for Advanced Studies), Mitaka, Tokyo 181-8588, Japan}

\author{Nobunari Kashikawa}
\affiliation{Department of Astronomy, School of Science, The University of Tokyo, 7-3-1 Hongo, Bunkyo-ku, Tokyo, JAPAN, 113-0033}
\affiliation{Optical and Infrared Astronomy Division, National Astronomical Observatory of Japan, Mitaka, Tokyo 181-8588, Japan}

\author{Roderik Overzier}
\affiliation{Observat\'orio Nacional, Rua Jos\'e Cristino, 77. CEP 20921-400, S\~ao Crist\'ov\~ao, Rio de Janeiro-RJ, Brazil}

\author{Jun Toshikawa}
\affiliation{
Institute for Cosmic Ray Research, The University of Tokyo, 5-1-5 Kashiwanoha, Kashiwa, Chiba 277-8582, Japan}

\author{Masafusa Onoue}
\affiliation{Max-Planck-Institut f\"ur Astronomie, K\"onigstuhl 17, D-69117 Heidelberg, Germany}


\author{Shogo Ishikawa}
\affiliation{Center for Computational Astrophysics, National Astronomical
Observatory of Japan, Mitaka, Tokyo 181-8588, Japan}
\affiliation{Division of Theoretical Astronomy, National Astronomical
Observatory of Japan, Mitaka, Tokyo 181-8588, Japan}


\author{Mariko Kubo}
\affiliation{Optical and Infrared Astronomy Division, National Astronomical Observatory of Japan, Mitaka, Tokyo 181-8588, Japan}

\author{Kei Ito}
\affiliation{Department of Astronomical Science, SOKENDAI (The Graduate University for Advanced Studies), Mitaka, Tokyo 181-8588, Japan}

\author{Shigeru Namiki}
\affiliation{Department of Astronomical Science, SOKENDAI (The Graduate University for Advanced Studies), Mitaka, Tokyo 181-8588, Japan}

\author{Yongming Liang}
\affiliation{Department of Astronomical Science, SOKENDAI (The Graduate University for Advanced Studies), Mitaka, Tokyo 181-8588, Japan}

\begin{abstract}
We have carried out deep and wide field imaging observations with narrow bands targeting 11 quasar fields to systematically study the possible 
photoevaporation effect of quasar radiation on surrounding low mass galaxies at $z\sim2-3$. 
We focused on Lyman alpha emitters (LAEs) at the same redshifts as the quasars that lie within the quasar proximity zones, 
where the ultra-violet radiation from the quasars is higher than the average background at that epoch. 
We found that LAEs with high rest-frame equivalent width of Ly$\alpha$ emission ($EW_0$) of $\gtrsim 150$\AA~with low stellar mass ($\lesssim 10^8 M_{\odot}$), are predominantly scarce in the quasar proximity zones, suggesting that quasar 
photoevaporation effects may be taking place. 
The halo mass of LAEs with $EW_0>150$\AA~is estimated to be $3.6^{+12.7}_{-2.3}\times10^9 M_{\odot}$ either from the Spectral Energy Distribution (SED) fitting or the main sequence. 
Based on a hydrodynamical simulation, the predicted delay in star formation under a local UV background intensity with 
$J (\nu_L)\gtrsim10^{-21}$ erg s$^{-1}$ cm$^{-2}$ Hz$^{-1}$ sr$^{-1}$ 
for galaxies having less than this halo mass is about $>20$ Myr, which is longer than the expected age of LAEs with $EW_0>150$\AA.
On the other hand,  the photoevaporation seems to be less effective around very luminous quasars, 
which is consistent with the idea that these quasars are still in an early stage of quasar activity. 
\end{abstract}


\section{INTRODUCTION}
High-$z$ luminous quasars are thought to form through major mergers of gas-rich galaxies 
\citep[e.g.][]{Kauffmann2002}. 
The typical bolometric luminosity of $\sim10^{46}$erg/s can be achieved by releasing the huge gravitational energy of gas falling toward a central super massive black hole (SMBH). 
It is, hence, speculated that quasars preferentially reside in overdense regions, where galaxy mergers happens frequently 
\citep[e.g.][]{Hopkins08}. 
The co-evolution of SMBHs and galaxies is supported by a tight correlation between the mass of central SMBHs and bulge stellar mass or velocity dispersion 
\citep[e.g.][]{Magorrian98, Marconi03}. 
Strong clustering of galaxies around quasars has been found beyond $z = 2$ \citep[e.g.][]{Shen07, Husband13, Morselli14, Garcia17}, while, some high-$z$ protoclusters without active galactic nuclei (AGN) activity have also been observed \citep[][]{Toshikawa14, Kang15}. 
Recently, \citet{Uchiyama2017} used wide field imaging of Hyper Suprime-Cam to statistically investigate the possible correlation between protoclusters and other overdensities and quasars at $z\!\sim\!4$. 
It was found that the most luminous quasars tend to avoid the overdense regions, which is supported by other studies  
 \citep[e.g.][]{Banados13, Mazzucchelli17, Kikuta17}. 
Recent MUSE (Multi Unit Spectroscopic Explorer) systematic observation also showed 
that quasar do not inhabit the most dense environments at $z\!\sim\!3$ and their environments seem to be similar to the field \citep[][]{Battaia18}. 
This is well explained by the fact that typical quasars occupy more average mass halos, but it has also been suggested that the most luminous quasars could suppress galaxy formation in their surroundings through feedback, even if some of these quasars reside in very massive dark matter halos. See \citet{Overzier16} for a more complete review. 

Photoevaporation could be effective around quasars, especially preventing the formation of low-luminosity galaxies. 
Low luminosity galaxies are closely bound to the surrounding inter-galactic medium (IGM). 
Photoionization heating by a strong ultra-violet (UV) background from quasars can evaporate the
collapsed gas in the halo and further inhibit gas cooling \citep[][]{Barkana1999}. 
This large-scale radiative feedback should be inefficient for bright ($L > L_{*}$) galaxies residing in deep potential wells, 
but may heavily suppress star formation in lower mass objects \citep[][]{Benson02a}. 
After the reionization epoch, low-luminosity galaxies can form in the general field only
if there are no strong UV ionizing sources such as quasars nearby, and consequently, 
it is expected that the luminosity function of galaxies around quasars is flatter than that of the general field population.

\citet{Kashikawa07} carried out a survey for both Lyman break galaxies (LBGs) and Lyman alpha emitters (LAEs) around a quasar, SDSS J0210-0018 at $z = 4.8$. 
They found that LBGs formed a filamentary structure including the quasar, while LAEs were distributed in a ring-like
structure around the quasar avoiding its immediate vicinity within a distance of $\sim 4.5 $ comoving Mpc (cMpc). 
This clustering segregation could be caused by photoevaporation resulting in a deficit of the lower mass LAEs around strong UV source. 
LAEs are young galaxies with low dust content, and relatively low stellar mass \citep[][]{Shapley01}; 
therefore, they are particularly prone to photoevaporation in the vicinity of strong radiation sources such as quasars.
Recently, \citet{Ota18} found a quasar residing in a low density region of LAEs at $z=6.6$,  
while \citet{Kikuta17} investigated distributions of LAEs around two quasars and a radio galaxy at $z\!\sim\!4.9$ 
and did not find any evidence of the quasar photoevaporation effect.   
These previous studies at $z \sim 5-6$ did not go deep enough to probe galaxies with sufficiently low luminosity (M$_{\rm{UV}} > -20.5$) or mass ($M_{\rm{vir}} < 10^{10} M_{\odot}$), where more effective photoevaporation is expected. 
In addition, the small sample sizes did not allow us to derive a general picture of this effect around high-$z$ quasars.

In this paper,  we carried out wide field imaging with Suprime-Cam mounted on the  8.2 m Subaru telescope 
targeting 11 fields with a strong local UV background from quasars at $z=2-3$ in order to systematically study the radiative feedback effect on low-luminosity galaxies. The combination of narrow-band (NB) imaging with deep broad-band (BB) imaging can effectively isolate  LAEs at $z=2-3$ \citep[][]{Venemans05}, down to low luminosity of $M^{*}_{\rm{UV}, z=3} + 2.0$, which is 1.5 magnitudes
 deeper than our previous study at $z = 4.8$.  

The paper is organized as follows. 
In \S 2, we describe the selection and reduction of targeted quasar fields, and construct our LAE sample. 
In \S 3, we investigate the LAE galaxy density in the vicinity of the quasars and search for correlations between the galaxy density and the properties of the LAEs and the quasars (e.g. black hole mass and luminosity).  
The implications of our results are discussed in \S 4.  
In \S 5 we conclude and summarize our findings. 
We assume the following cosmological parameters: 
$\Omega_{M} = 0.3 $, $\Omega_{\Lambda} = 0.7$, $H_{0} = 70~ $km s$^{-1}$ Mpc$^{-1}$. 
Magnitudes are given in the AB system. 

\begin{figure}
\begin{center}
 \plotone{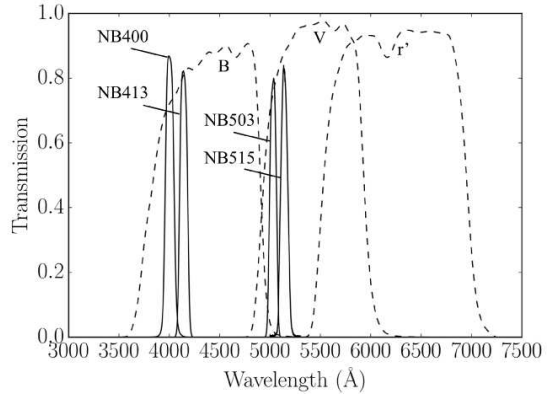}
\end{center}
\caption{Filter transmission curves. The solid lines show the transmission curves of broad-band filters, $B$, $V$, $r'$, from left. 
The dashed lines indicate the curves of narrow-band filters, $NB400$, $NB413$, $NB503$ $NB515$, from left. 
 }\label{filter}
\end{figure}


\begin{deluxetable*}{lllllllll}[t!]
\tablecaption{Taget quasar sample \label{t1}}
\tablecolumns{14}
\tablenum{1}
\tablewidth{0pt}
\tablehead{
\colhead{Quasar Name} & \colhead{R.A. (J2000)} & \colhead{Decl. (J2000)} & \colhead{$z_{\text{sys}}$\tablenotemark{a}} & \colhead{log$\lambda L_\lambda$ \tablenotemark{b}} & \colhead{ log$ M_{BH}$ \tablenotemark{c}}& \colhead{ radio \tablenotemark{d}}& \colhead{ ref. 1 \tablenotemark{e}}& \colhead{ ref. 2 \tablenotemark{f}}\\
\colhead{} & \colhead{} & \colhead{} & \colhead{} & \colhead{(erg s$^{-1}$)}& \colhead{($M_{\odot}$)}& \colhead{}& \colhead{}& \colhead{}
}
\startdata
SDSS J095141.33+013259.5   &  $09^{h}51^{m}41.33^{s}$ & $+01^{\circ}32'59.5'' $ &  2.411          &      45.9534           &  $8.90^{+0.20}_{-0.20}$ & quiet    &  1 &1            \\ 
OH91--121                                 & $12^{h}35^{m}02.6^{s}$  & $-11^{\circ}30'29'' $       & 2.407          &      46.9242     	 &  $9.89^{+0.05}_{-0.05}$  &   quiet   & $*$  & 2       \\ 
SDSS J125034.41--010510.5   &  $12^{h}50^{m}34.41^{s}$  & $-01^{\circ}05'10.5''$  &  2.397        &    45.8869           &  $9.06^{+0.20}_{-0.20}$  &  quiet        &  1   & 1          \\ 
SDSS J141123.51+004253.0   & $14^{h}11^{m}23.5^{s}$    & $+00^{\circ}42'53'' $       & 2.257         &     46.7248   	 &  $9.37^{+0.30}_{-0.30}$  &  loud   &  3  & 3           \\ 
SDSS J142656.18+602550.8           & $14^{h}26^{m}56.1^{s}$  & $+60^{\circ}25'50'' $ & 3.202        &       47.8206      &  $9.82^{+0.30}_{-0.30}$   &  quiet   & 3     & 3            \\ 
TXS 1529--230                          & $15^{h}32^{m}31.5^{s}$  & $-23^{\circ}10'32'' $       & 2.280        &       46.3152       &  $9.42^{+0.13}_{-0.13}$   &     loud  &  $*$   & 2      \\ 
SDSS J155137.22+321307.5   & $15^{h}51^{m}37.22^{s}$  & $+32^{\circ}13'07.5'' $  & 3.143         &        46.0280     	 &  $9.54^{+0.18}_{-0.18}$  &   quiet  &  4  & 4     \\ 
SDSS J162359.21+554108.7   & $16^{h}23^{m}59.21^{s}$  & $+55^{\circ}41'08.7'' $ & 2.272           &      45.5920      &  $8.86^{+0.05}_{-0.05}$   &  quiet &  $*$  & 2       \\  	      
SDSS J162421.29+554243.0   & $16^{h}24^{m}21.29^{s}$  & $+55^{\circ}42'43.0'' $ & 2.278      &      45.6695        &   $8.59^{+0.04}_{-0.04}$      &    quiet   &  $*$ & 2      \\ 
SDSS J170102.18+612301.0   &$17^{h}01^{m}02.18^{s}$   & $+61^{\circ}23'01.0'' $ & 2.301         &        46.4573       &  $9.72^{+0.30}_{-0.30}$    &  quiet   & 3 & 3             \\ 
HB89--1835+509                     & $18^{h}36^{m}14.50^{s}$  & $+51^{\circ}01'45.0'' $  & 2.272         &       46.4891      &  $9.34^{+0.06}_{-0.06}$  &   quiet  &  $*$    & 2   \\ 
SDSS J213510.60+013930.5   & $21^{h}35^{m}10.6^{s}$  & $+01^{\circ}39'31'' $     & 3.199        &       46.2128       &  $8.65^{+0.30}_{-0.30}$    &  loud &  3   & 3    
\enddata
\tablenotetext{a}{Systemic redshift. }
\tablenotetext{b}{Logarithm of intrinsic luminosity at rest-wavelength 912\AA~in $L_\odot$, which is estimated by fitting a single power law to the continuum free of line emission (\S3.1). }
\tablenotetext{c}{Logarithm of black hole mass in $M_{\odot}$. }
\tablenotetext{d}{Radio loud or not \citep[][]{Shemmer04, Sulentic14}. }
\tablenotetext{e}{The reference for the black hole mass. $*$ shows that the black hole mass was estimated by us using the CIV-based black hole mass estimator of \citet{Shen11} using the parameters given by \citet{Sulentic14}. \citet{Netzer07} found that the errors of the black hole masses are estimated to be $20-60$\%. We provided the error of 60\% for their black hole masses. The typical uncertainties of the black hole masses estimated by \citet{Shemmer04}  are less than a factor of two. Their masses we list have the uncertainty of the factor of two. }
\tablenotetext{f}{The reference for $z_{\text{sys}}$}
\tablecomments{ (1) \citet{Netzer07}; (2) \citet{Sulentic14}; (3)  \citet{Shemmer04}; (4)   \citet{Saito16} }
\end{deluxetable*}

\section{DATA AND SAMPLE SELECTION}
\subsection{Observations and Reduction}
We conducted wide field imaging with Suprime-Cam \citep[][]{Miyazaki02} on the 8.2m Subaru telescope for 11 quasar fields at $z\!\sim\!2-3$ in order to systematically study the photoevaporation effect. 
The quasar fields were selected based on the following criteria. 
We selected quasars, whose redshifted Ly$\alpha$ falls in the wavelength range of the Suprime-Cam NB filters, $NB400$ ($\lambda_{\rm{c}} =4000$\AA, FWHM$=92$\AA, $z_{\rm{Ly\alpha}}=2.29_{-0.04}^{+0.04}$), $NB413$ ($\lambda_{\rm{c}}=4130$\AA, FWHM$=83$\AA, $z_{\rm{Ly\alpha}}=2.40_{-0.04}^{+0.03}$), $NB503$  ($\lambda_{\rm{c}}=5030$\AA, FWHM$=74$\AA, $z_{\rm{Ly\alpha}}=3.14_{-0.04}^{+0.03}$) or $NB515$ ($\lambda_{\rm{c}}=5150$\AA, FWHM$=79$\AA, $z_{\rm{Ly\alpha}}=3.24_{-0.04}^{+0.03}$), 
 where $z_{\rm{Ly\alpha}}$ shows the Ly$\alpha$ redshift ranges corresponding to the FWHMs of the NB filters.
The systemic redshifts are determined by H$\beta$, [OIII] emission lines 
\citep[][]{Netzer07, Shemmer04} or narrow low ionization lines \citep[][]{Sulentic14}, 
which are more reliable than those measured by Ly$\alpha$ or CIV lines which are easily influenced by quasar outflows 
\citep[][]{Uchiyama2017}. 
We selected quasars with a wide range in UV luminosity (log$\lambda L_\lambda (912$\AA )$=45.6-47.8$)
in order to investigate the variation of the quasar photoionization feedback. 
In total, we observed 11 quasar fields of which three fields contain radio-loud quasars (SDSS1411, TXS1529 and SDSS2135) and one field contains a quasar-pair with two quasars (SDSS1623 and SDSS1624) in close proximity to each other \citep[e.g.][]{Djorgovski87, Onoue17a}.  
The redshift and angular separation of the quasar-pair are $\Delta z=0.006$ and $\Delta \theta = 3.5$ arcmin ($1.7$ pMpc), respectively. 
The properties of the quasars are also summarized in Table \ref{t1}.  

Observations were made on six nights in UT $2014$, $2015$ and $2017$. 
Suprime-Cam has ten $2$k $\times 4$k MIT/LL CCDs, and covers a contiguous area of $34'\times27'$ with a pixel scale of $0.''202$ pixel$^{-1}$.
BB imaging is required to measure the flux excess in NB filters. 
The integration time was $12000$--$16800$ s in each band. 
For the BB filters, the typical unit exposure time was $600$ s for $B$-band and $V$-band filters and $300$ s for the  $r'$-band filter, 
which was used because the $V$-band filter was not available during the observations of the SDSS1551 quasar field, 
while for the NB filters, the exposure time was $1200$ s. 
We adopted a common circle dithering pattern (full cycle) consisting of $7$--$9$ pointings for NB filters and $4$--$8$ pointings for BB filters. 
The sky condition was fairly good with a seeing size of $0.60$--$1.40$ arcsec. 
The observation configuration is shown in Table \ref{t2}.


\begin{deluxetable*}{llllll}[t!]
\tablecaption{Observation log \label{t2}}
\tablecolumns{24}
\tablenum{2}
\tablewidth{0pt}
\tablehead{
\colhead{Quasar Name} & \colhead{filter} & \colhead{exposure time $\times$ shots} & \colhead{seeing} & \colhead{$m_{lim, 5\sigma}$} & \colhead{Date}\\
\colhead{} & \colhead{} & \colhead{(sec)} & \colhead{(arcsec)} & \colhead{(mag)}& \colhead{}
}
\startdata
SDSS J095141.33+013259.5   &  $NB400$ &  $1200\times9$    &  $1.''24   $    &  25.23          &   2014 May 25, 26, 27   \\ 
                                                 &    $B$        &   $600\times5$   &  $0.''96    $     &  26.44        &     2014 May 27          \\ \hline
OH91--121                                 & $NB413$  &  $1200\times7$     &  $1.''20 $     &  25.45         &         2015 Jun 17         \\ 
                                                 &    $B$         &  $600\times7$     & $1.''20  $      &  26.23         &        2015 Jun 16       \\ \hline
SDSS J125034.41--010510.5   &  $NB413$  &  $1200\times8$     &  $1.''40 $    &  25.77          &    2014 May 26, 27     \\ 
                                                 &   $B$          & $300\times8$  & $1.''40 $     &  26.12          &    2014 May 26            \\ \hline
SDSS J141123.51+004253.0   & $NB400$    & $1200\times8$    & $1.''04 $       &  25.75          &     2014 May 25                 \\ 
                                                 &      $B$       &  $600\times5$   & $1.''30 $      &  26.25          &      2014 May 26        \\ \hline
SDSS J142656.18+602550.8   & $NB515$ &  $1200\times7$  & $0.''78  $     &  25.63         &       2014 May 27           \\ 
                                                 &   $V$        &  $600\times4$     &  $0.''80 $     &  25.95         &       2014 May 27            \\ \hline
TXS 1529-230                          & $NB400$   &  $1200\times5$   & $0.''96$       &  25.00          &       2015 Jun 16        \\ 
                                                 & $B$           &   $600\times4$  &  $0.''96 $      &  26.04           &        2015 Jun 17       \\ \hline
SDSS J155137.22+321307.5   & $NB503$   & $900\times4+600\times10$   & $0.''60 $       &  25.28          &        2017 May 23                \\ 
                                                 &    $r'$  \tablenotemark{a} & $300\times6$                       &  $1.''12 $         &  25.80                &          2017 May 23       \\ \hline
SDSS J162359.21+554108.7   & $NB400$   & $1200\times7$   & $1.''20 $          &  25.67           &      2015 Jun 17      \\ 
SDSS J162421.29+554243.0   &   $B$          &  $600\times4$ &   $0.''98  $       &  26.88          &       2015 Jun 17          \\ 	   \hline    
SDSS J170102.18+612301.0   & $NB400$   &    $1200\times9$   & $1.''10 $     &  25.72         &   2014 May 25                   \\
                                                 &      $B$      &    $600\times4$    &  $0.''70 $   &   27.12         &  2014 May 27                    \\ \hline
HB89--1835+509                     & $NB400$   &  $1200\times8$   & $0.''84  $      &  25.73         &       2015 Jun 16           \\ 
                                                 &    $B$         &  $600\times4$  &  $0.''94  $      &  26.59      &        2015 Jun 17               \\ \hline
SDSS J213510.60+013930.5  & $NB515$ &    $1200\times8$  & $1.''10 $    &  24.95          &       2014 May 26         \\ 
                                                 &    $V$       &    $600\times4$   &  $0.''80$     &  25.48        &       2014 May 27       \\ \hline
\enddata
\tablenotetext{a}{
The $r'$-band was used because the $V$-band filter was not available for SDSS1551.
}
\end{deluxetable*}

We used SDFRED version 2.0 \citep[][]{Yagi02, Ouchi04} to conduct the data reduction of NB and BB images. 
The L.A. Cosmic recipe \citep[][]{vanDokkum01} was used to remove cosmic rays, which was significant in the NB images taken with a long exposure time ($1200$ s).
Flux calibrations of NB images were made with spectro-photometric standard star 
Feige110 for $NB400$, $NB413$ and $NB515$ images and Feige63 for $NB503$ images. 
For BB images, we used photometric standard stars SA110 and SA113 for BB images, and SA110 for V-band images. 
The calibrations for $r'$-band were done using stars in the Sloan Digital Sky Survey (SDSS) DR14 catalog. 
We carefully checked the estimated zero-point magnitudes using spectroscopic A-type stars, 
which have an almost flat spectrum in our observed wavelength ranges, found in SDSS DR14.  
The astrometric calibration of NB-detected objects was performed by using the Naval Observatory Merged Astrometric Dataset (NOMAD). 
The astrometric accuracy is $\sim0.3$ arcsec estimated by using the SDSS DR14 catalog. 
The double image mode of SExtractor version 2.19.5 \citep[][]{Bertin96}  was used for object detection and photometry. 
We detected objects that had at least $10$ connected pixels above $1.5\sigma$ times the sky background rms noise and made photometric measurements at the $1.5\sigma$ level in NB images. 
The sky background rms was estimated through SExtractor as follows. 
Each image was separated into several meshes of 24 arcsec. 
The mean and standard deviation of the counts in each mesh were estimated with $3\sigma$ clipping, and 
the mesh values which were too bright were removed through the median filtering.  
Finally, each mesh has the sky background rms interpolated by a bicubic interpolation. 
We masked the area around bright objects and shallower regions around the edge of each image ($\sim3$ arcmin). 
The magnitudes of detected objects were measured within apertures of $2$ arcsec diameter. 
The $5\sigma$ limiting magnitudes measured in $2$ arcsec apertures are summarized in Table \ref{t2} for all images.  

\begin{figure*}
 \plotone{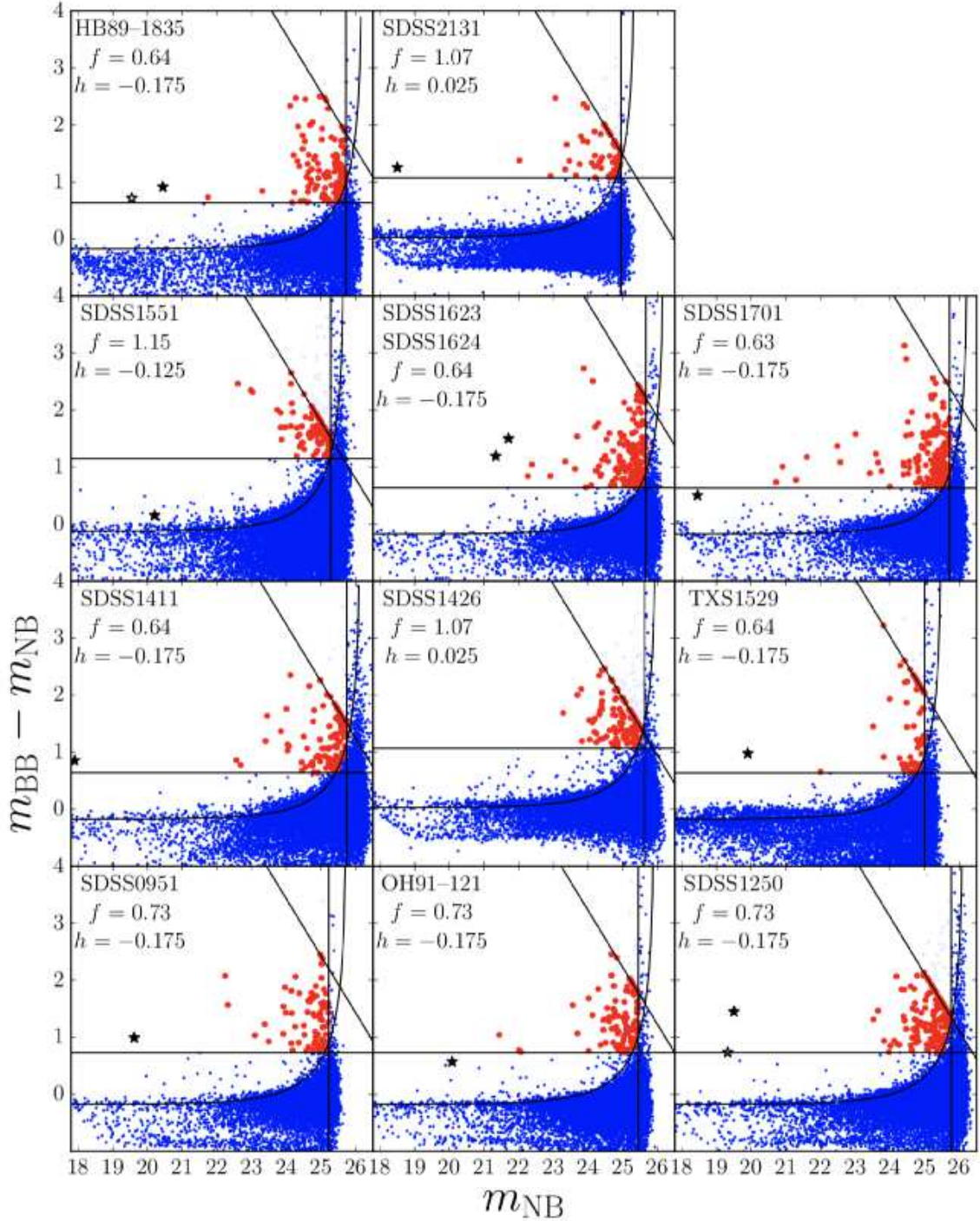}
\caption{Color magnitude diagram $m_{\rm{BB}}-m_{\rm{NB}}$ versus $m_{\rm{NB}}$. The blue points show all objects except those in the masked regions around bright stars. The red points indicate the LAE candidates and the black stars show the quasars in each field. The horizontal lines correspond to $EW_0=20$\AA. The $5\sigma$ NB limiting magnitudes in each field are drawn by the vertical lines and the $h+3\sigma$ color errors are shown by the curved black lines. 
The diagonal lines show the $2\sigma$ BB limiting magnitudes in each quasar field.  
There is no excess in the color of the SDSS1551, SDSS1701 and OH91--121 quasars due to strong absorption seen in the spectrum at the Lyman $\alpha$ line. In addition, the flux of quasar SDSS1426 could not be accurately measured due to saturation. One field contains a quasar-pair (SDSS1623 and SDSS1624) in close proximity to each other.  
\label{colormag}}
\end{figure*}
 
\begin{figure*}
\begin{center} 
 \plotone{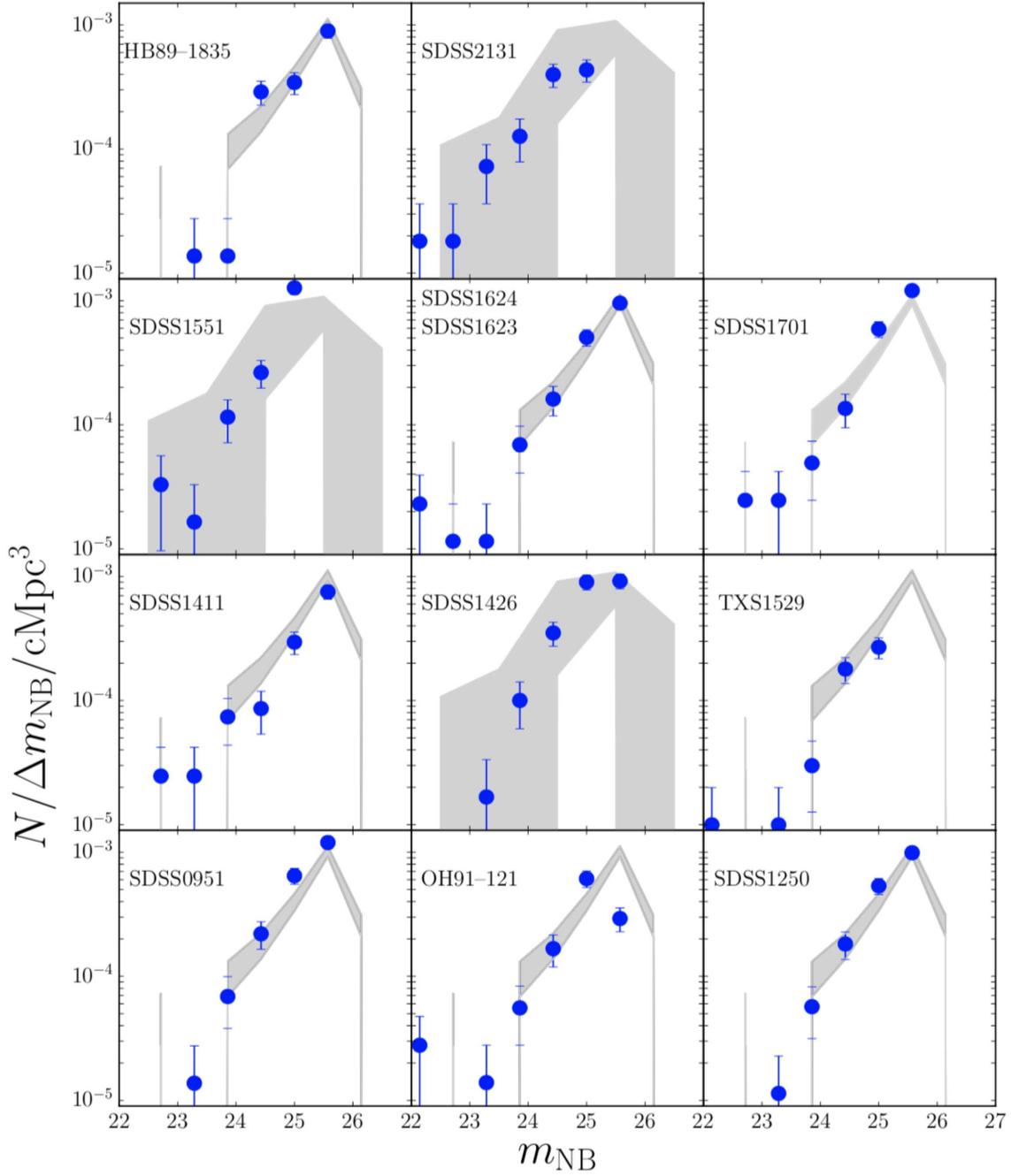}
\end{center} 
\caption{Number count of LAEs. The blue dots show the number counts of LAEs in each quasar field. 
The error bars indicate the Poisson error. The gray shaded region shown in the HB89--1835, SDSS1623+1624, SDSS1701, SDSS1411, TXS1529, SDSS0951, OH91--121, and SDSS1250 quasar fields at $z\sim2$ indicates the LAE number count of \citet{Mawatari12}, while in the SDSS1551, SDSS1426, and SDSS2131 fields at $z\sim3$ they indicate the count given in \citet{Grove09}. }\label{count}
\end{figure*}

\subsection{LAE Selection} 
We used the following criteria, which are essentially the same as \citet{Mawatari12} and \citet{Matsuda05}, to select LAEs in all quasar fields :
\begin{eqnarray}
  m_{\rm{BB}}-m_{\rm{NB}}  & > & f(EW_0=20\rm{\AA}),  \label{eq1}\\ 
          m_{\rm{NB}} &  < & m_{\rm{lim},5\sigma}, \\
  m_{\rm{BB}}-m_{\rm{NB}} & > & h + 3\sigma_{\rm{color}}, 
\end{eqnarray}
where $m_{\mathrm{lim}, 5\sigma}$ and $3\sigma_{\rm{color}}$ are the $5\sigma$ limiting magnitude and $3\sigma$ $m_{\rm{BB}}-m_{\rm{NB}}$ color error, respectively. 
$h$ is the color term, for which we use the typical $m_{\rm{BB}}-m_{\rm{NB}}$  color of the galaxies without Ly$\alpha$ emission, assumed to be the mode in the color distribution of objects lying in a range of $18.0<m_{\rm{NB}}<24.0$. $f(EW_0=20$\AA$)$ is the color which corresponds to the Ly$\alpha$ equivalent width ($EW_0$) of $20$\AA~at rest frame, expected by our LAE model constructed in Appendix A. 
If the objects were not detected in the BB filter at 2$\sigma$, their BB magnitudes were replaced by the corresponding 2$\sigma$ limiting magnitudes.  
In addition, we carefully checked the images and removed fake detections by eye. 
As a result, we obtained 1171 LAEs in total.  195 of them had no UV continuum detection.  
These LAEs may appear due to the quasar fluorescence effect \citep[e.g.][]{Cantalupo07}.  
It was confirmed that even if the 195 LAEs are excluded, our results did not change within the 1$\sigma$ error. 
The number of LAE candidates obtained through the criteria in each field is also listed in Table \ref{t2} and the color magnitude diagrams of each field are shown in Figure \ref{colormag}. 
Figure \ref{count}  shows the number counts of LAEs in each quasar field. 
The counts are consistent with those of previous studies \citep[][]{Mawatari12, Grove09}, who studied a field around a radio galaxy 53W002 (744 arcmin$^2$) and some blank fields BRI 1202--0725, BRI 1346--0322 and Q 2138--4427 (133 arcmin$^2$), respectively.  
We can ignore the contamination of low-$z$ emitters because the contamination rate is expected to be at most $\sim 1$\% according to \citet{Mawatari12, Matsuda05, Matsuda06}. 
There is no excess in the $m_{\rm{BB}}-m_{\rm{NB}}$ color of the SDSS1551, SDSS1701 and OH91--121 quasars due to strong absorption seen in the spectrum at the Lyman $\alpha$ line. In addition, the flux of quasar SDSS1426, which is not plotted in the color magnitude diagram, could not be accurately measured due to  saturation of the CCD. 
Two possible quasars, SDSS J125036.70-005531.7 \citep[][]{Lee13} and QSO B1833+509 \citep[][]{Weedman85}, whose redshifts determined by Ly$\alpha$ emission coincidentally fall in the NB filters reside in the SDSS1250 and OH91--121 quasar fields, also shown in Figure \ref{colormag} (black open stars), respectively.  
We confirmed that even if the two quasars are included in our analysis, the results did not change.  



The detection completeness of LAEs was estimated by Monte Carlo simulation.
First, we distribute 5000 artificial objects with $18.0<m_{\rm{NB}}<m_{\rm{lim},5\sigma}$ 
so that the EW distribution follows that of \citet{Mawatari12}, $f(EW_0)=Ce^{-EW_0/w_0}$, where $C$ is a normalization constant and $w_0=43.7$ \AA~is the $e$-folding length. 
We conduct source detection, photometry, and LAE selection with the same parameters as we did for the real images. 
Then, 
we define the completeness as the fraction of the number of pseudo LAEs detected in each magnitude bin (we take an ensemble average with 1000 realizations).  
The completeness is $\gtrsim60$\% at the 5$\sigma$ limiting magnitude of the NB filter for each field.


\begin{figure*}
\begin{center}
 \plotone{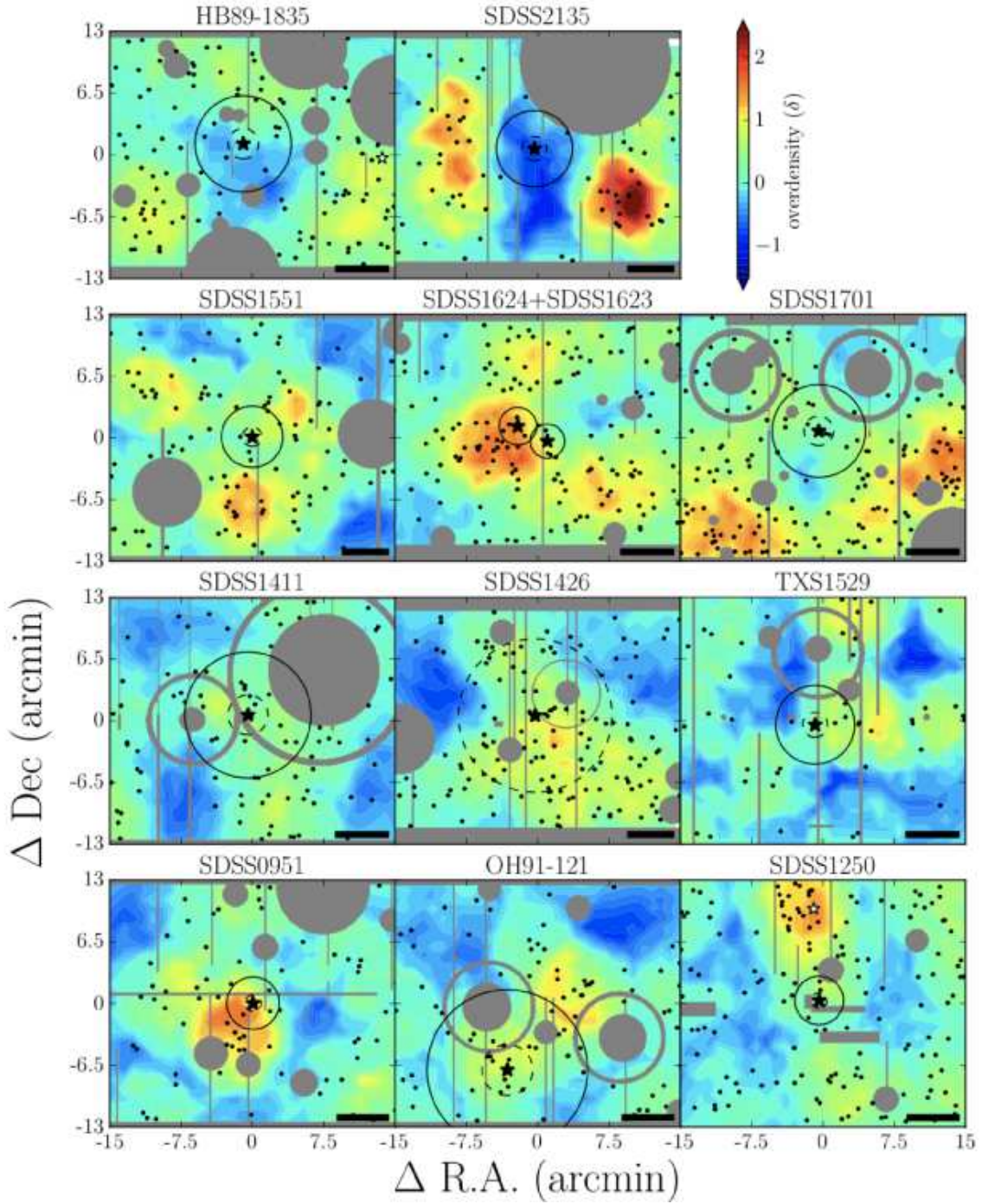}
\end{center}
\caption{Overdensity maps in the quasar fields. The filled and open black stars show the quasars and possible quasars in each field, respectively. 
The black dots indicate the LAE candidates. 
The color contours show the overdensity. The radii of isotropic UV intensity $J_{21}$ of $1$ and $10$ are indicated by the solid and dashed black circles, respectively. 
The gray shaded regions show the masked regions. The size of each panel is $ 30 \times 25 $ arcmin$^2$. 
The length of 8 cMpc, which is the scale used to derive the overdensity,  is indicated by the black line in the lower right corner in each panel.  }\label{density}
\end{figure*}

\begin{figure}
\begin{center} 
 \plotone{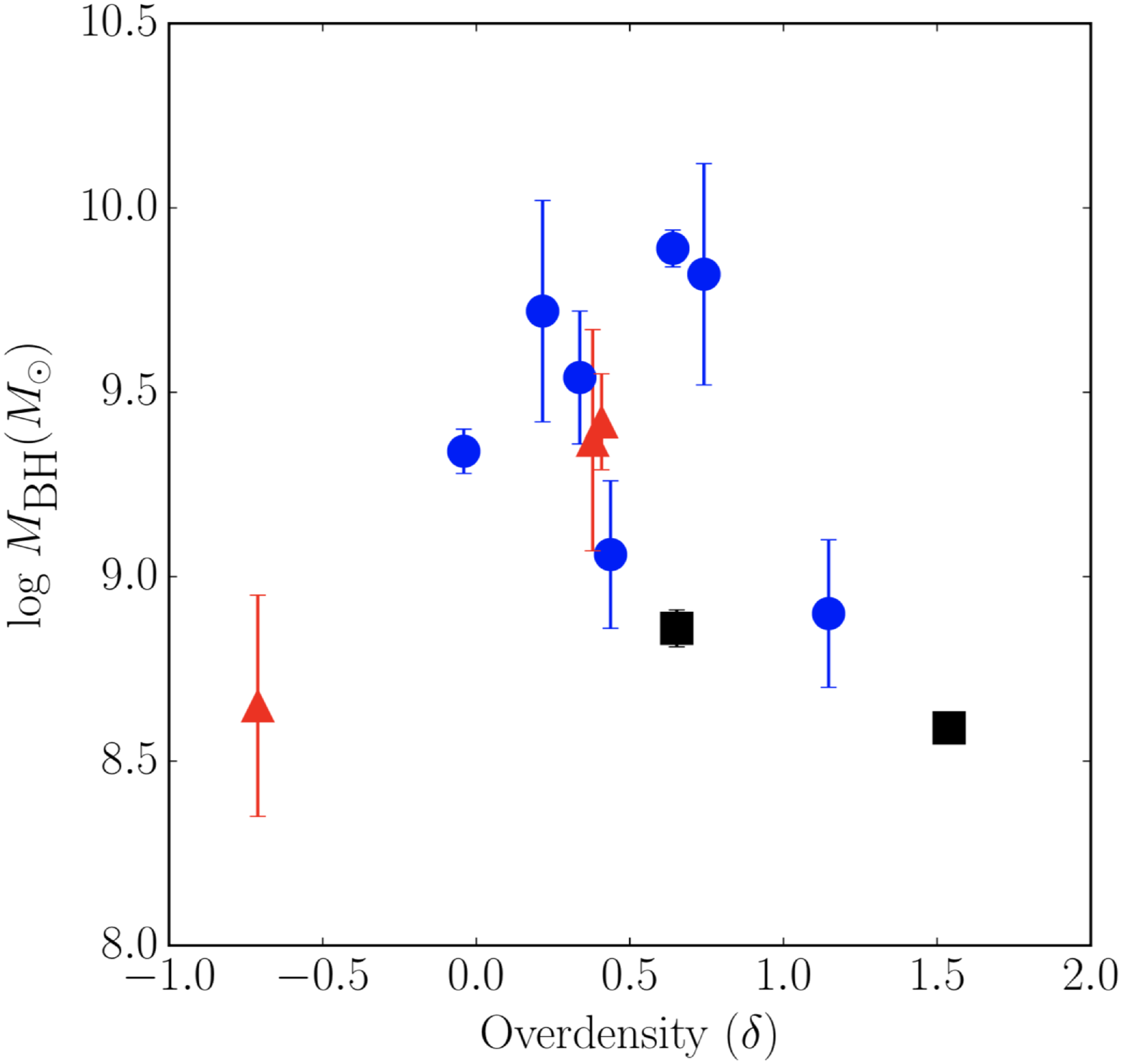}
\end{center}
\caption{Relation between black hole mass and overdensity. The red triangle and blue points show the radio-loud and radio-quiet quasars, respectively.  
The quasar pair is indicated by black squares. 
}\label{prop}
\end{figure}

\section{RESULTS}
\subsection{LAE Galaxy Density around Quasars}
We applied the fixed aperture method to determine LAE surface density contour maps. 
We estimated a local number density by counting LAEs within the fixed aperture to evaluate the overdensity $\delta$ defined as $\delta = (N-\bar{N})/\bar{N}$, where $N$ is the number of LAEs in the aperture. 
The scale of $8.0$ cMpc is used as the radius of distributed apertures in each field \citep[][]{Kikuta17}.  
 $\bar{N}$ is the average of $N$ given by the all-quasar-fields average of the LAE number counts at the $5\sigma$ NB limiting magnitude for each field.
We confirmed that the LAE number counts are consistent each other for all our target regions. 
The surface density of LAEs in masked regions is assumed to be the same as $\bar{N}$. 

We also calculate the size of the quasar proximity zones, where the UV radiation from the quasars is higher than the UV background radiation. 
The isotropic UV intensity $J(\nu)$ of radiation at the Lyman limit from a central quasar, $J_{21}$, is important to evaluate the quasar photoevaporation effect \citep[][]{Kashikawa07}:  
\begin{equation}
J(\nu) = J_{21} \left( \frac{\nu}{\nu_L} \right)^\alpha \times 10^{-21} ~ \rm{erg}~\rm{cm}^{-2}~ \rm{s}^{-1}~ \rm{Hz}^{-1}~ \rm{sr}^{-1} ,  
\end{equation}
where $\nu_L$ is frequency at the Lyman limit, and $\alpha$ is the slope of the flux density of the quasar, $F_{\nu} (\nu) \propto \nu^\alpha$. 
We estimate the intrinsic monochromatic luminosity at the Lyman limit by fitting a single power law to the flux free of line emission at $1340$---$1360$ \AA, $1440$---$1450$\AA~and $1700$---$1730$\AA, using the rest-frame wavelength quasar spectrum
from the SDSS Science Archive Server (SAS), 
with spectral resolution $R\sim1300$---$2500$ \citep[][]{Paris17}, for the quasars obtained by SDSS. 
The spectra of OH91--121, TXS1529, and HB89--1835 were taken from \citet{Sulentic14}.  
The intensity of the UV background radiation at $z=2-4.5$ was evaluated to be $J_{21}=1.0_{-0.3}^{+0.5}$ with weak dependency of redshift by the QSO proximity effect measurements of \citet{Cooke97}.  
The local UV radiation within the circle of the radius of $r_{\rm{prox}}$, 
\begin{equation}
r_{\rm{prox}} = \frac{1}{4 \pi}\sqrt{\frac{L(\nu_L)}{10^{-21}}} ~~\rm{pMpc},  \label{eq:1} 
\end{equation}
is expected to be enhanced compared with the UV background. 
Therefore, we use $r_{\rm{prox}}$ to search for possible photoevaporation effects on the properties of LAEs.  
For SDSS1426, we used the $J_{21}=10$ radius instead of the $J_{21}=1.0$ to evaluate the environmental effect 
because the radius of $J_{21}=1.0$ is larger than the FoV due to the brightness of the quasar. 
We estimated the $r_{prox}$ for each quasar using its own alpha. 
Assuming the fixed typical alpha of $-0.30$  \citep[][]{Selsing16}, 
the $r_{\text{prox}}$ varies 
by as little as $\sim0.04$arcmin ($\sim0.07$cMpc). 
The values of $r_{\rm{prox}}$ are listed in Table \ref{t3}. 
Figure \ref{density} shows the density maps of the LAEs in each quasar field with $r_{\rm{prox}}$ indicated for $J_{21}=1$ and $10$. 
 

\begin{deluxetable*}{lllllllll}[h]
\vspace{-2mm}
\tablecaption{Results \label{t3}}
\tablecolumns{14}
\tablenum{3}
\tablewidth{0pt}
\tablehead{
\colhead{Quasar Name} & \colhead{volume\tablenotemark{a}} & \colhead{\# LAEs \tablenotemark{b}} & \colhead{$\delta$ \tablenotemark{c}} & \colhead{$r_{\rm{prox}}$ \tablenotemark{d}} & \colhead{ $N^{\rm{exp}}_{<r_{\rm{prox}}}$\tablenotemark{e}}& \colhead{$N^{\rm{obs}}_{<r_{\rm{prox}}}$ \tablenotemark{f}}\\
\colhead{} & \colhead{($\times10^{5}$cMpc$^3$)} & \colhead{} & \colhead{} & \colhead{(arcmin (cMpc))}& \colhead{}& \colhead{}
}
\startdata
SDSS J095141.33+013259.5   & 1.27                                  & 68                &  1.15        &  $2.76$ ($4.60$)                    &  0.368  & 0.00  \\ 
OH91--121                                 & 1.26                                  & 84                 & 0.640       &  $8.44$ ($14.0$)                   &  4.22    &  3.63\\ 
SDSS J125034.41-010510.5   &  1.54                                  & 157              &  0.437       &    $2.56$ ($4.24$)                  & 0.464   & 0.00  \\ 
SDSS J141123.51+004253.0    & 1.42                                  & 100               & 0.379       &  $6.63$ ($10.7$)                  &  2.17  & 2.52 \\ 
SDSS J142656.18+602550.8     & 1.05                               & 137              & 0.741       &  $25.5$ ($48.6$)                   & 0.866\tablenotemark{g}   &  7.45\tablenotemark{g}\\ 
TXS 1529-230                          & 1.76                                   & 48                & 0.408       &  $4.15$ ($6.70$)                    &  0.316   &  0.00 \\ 
SDSS J155137.22+321307.5   & 1.06                                  & 103              & 0.337       &  $3.22$ ($6.08$)                     &  0.00  &  0.00\\ 
SDSS J162359.21+554108.7   & 1.52                                  & 142              & 0.653        &  $1.80$ ($2.91$)                   &   0.188   & 0.00 \\  	      
SDSS J162421.29+554243.0   & 1.52                                  & 142              & 1.54          &  $1.97$ ($3.19$)                   &   0.253  &  0.00\\ 
SDSS J170102.18+612301.0   &1.42                                   & 162              & 0.216       &  $4.89$ ($7.95$)                    & 12.8  &  0.00\\ 
HB89--1835 +509                     & 1.27                                  & 111               & -0.0406  &  $5.06$ ($8.17$)                    &  2.46   &  1.00\\
SDSS J213510.60+013930.5    & 0.967                                & 59                & -0.711     &  $4.01$ ($7.63$)                   & 0.163    &  0.00 \\   \hline 
Total\tablenotemark{h}                       &                                            &             &                 &                                                    &    23.4      &  7.15 \\
Total (exc. brightest quasars)\tablenotemark{i} &                  &            &                  &                                                          &     14.6       & 0.00 
\enddata
\vspace{-2mm}
\tablenotetext{a}{Effective volume in each field. }
\vspace{-2mm}
\tablenotetext{b}{The observed number of LAEs. }
\vspace{-2mm}
\tablenotetext{c}{Overdensity which quasar reside in. $\delta = (N-\bar{N})/\bar{N}$. }
\vspace{-2mm}
\tablenotetext{d}{The local UV radiation within the circle of the radius of $r_{\rm{prox}}$, which is defined by equation (\ref{eq:1}), is enhanced compared with the UV background by the radiation from each quasar. }
\vspace{-2mm}
\tablenotetext{e}{The expected completeness-corrected number of LAEs with $EW_0>150$\AA~  in each quasar proximity zone. }
\vspace{-2mm}
\tablenotetext{f}{The observed completeness-corrected number of LAEs with $EW_0>150$\AA~ in each quasar proximity zone. }
\vspace{-2mm}
\tablenotetext{g}{ We used the $J_{21}=10$ radius instead of the $J_{21}=1.0$ radius, which is used in other quasar fields, 
because the radius of $J_{21}=1.0$ in the SDSS1426 field is larger than its FoV. } 
\vspace{-2mm}
\tablenotetext{h}{ Total but for the SDSS1426. }
\vspace{-2mm}
\tablenotetext{i}{Total but for the four brightest quasars, SDSS1411, SDSS1426, OH91--121 and HB89--1835. }
\end{deluxetable*}

The average density near the quasars is 0.023 cMpc$^{-2}$, which corresponds to an overdensity $\delta=0.48$ with $0.82\sigma$ significance. 
The quasars are found to reside in average density environments, which is consistent with \citet{Uchiyama2017}. 
SDSS1624 occupies the highest density of 0.045 cMpc$^{-2}$ which corresponds to overdensity $\delta =1.539$.  
Interestingly, this quasar forms a close pair with another quasar.  
\citet{Onoue17a} found that quasar pairs statistically tend to reside in overdense regions at $z\!\sim\!1$ and $4$. 
The densities of the quasars are summarized in Table \ref{t3}. 
We examine possible relations between the overdensities and black hole masses and radio loudness of the quasars, shown in Figure \ref{prop}. 
There are no significant correlation as in the case of \citet{Uchiyama2017}. 
In the Spearmann rank correlation test, the $P$-value of a relation between the overdensity and the black hole mass is 0.17. 
Using only radio-quiet quasars, the $P$-value is 0.24. 


\subsection{UV Luminosity Distributions of LAEs in the Vicinity of Quasar}
\begin{figure*}
\begin{center}
 \plotone{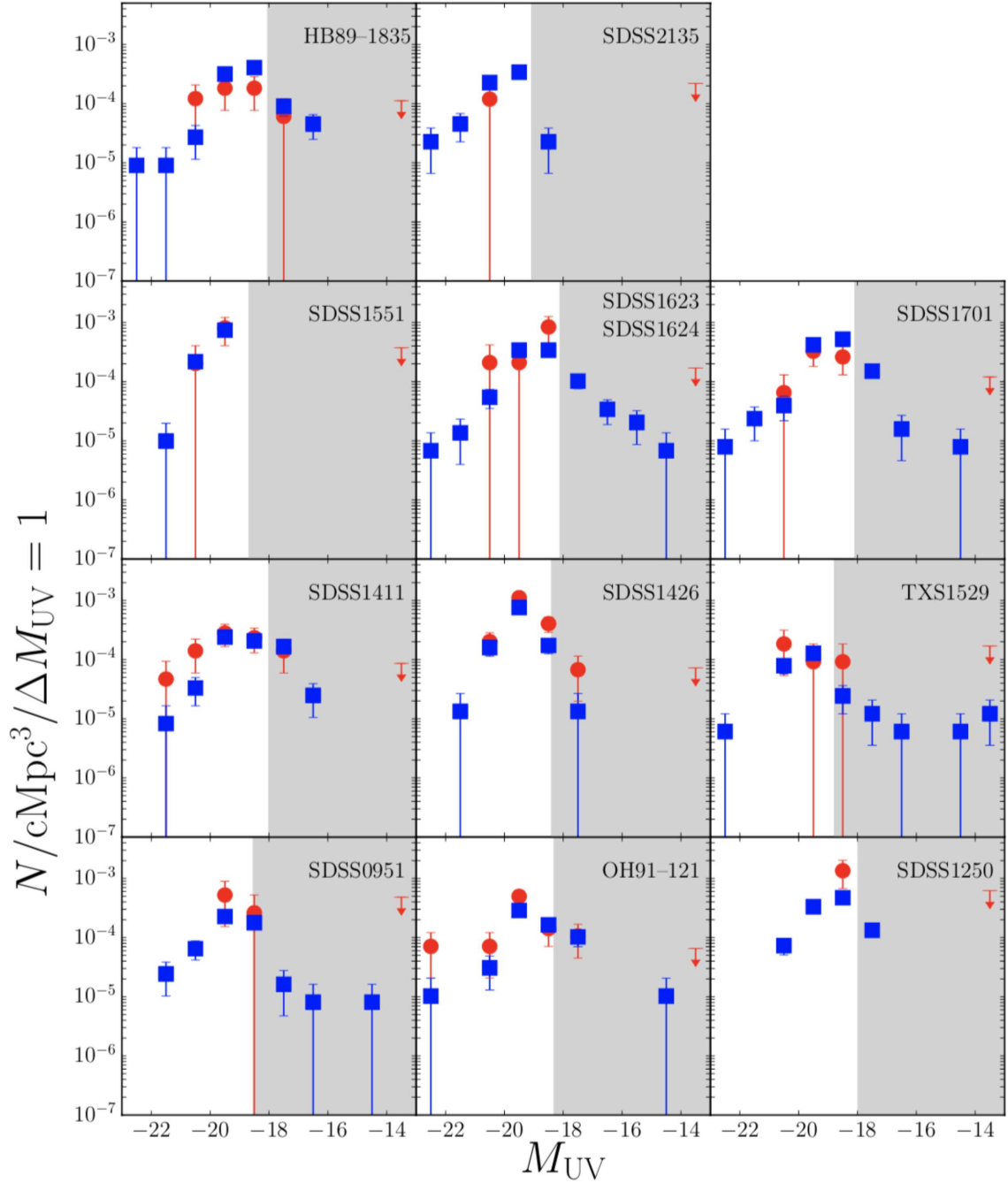}
\end{center}
\caption{Distribution of $M_{\rm{UV}}$ of LAEs. 
The red circles and blue squares show the UV luminosity distribution  in the vicinity and outer region of quasars, respectively.  
The error bar assumes a binomial distribution \citep[][]{Gehrels86}. The gray shade region in each field shows the incomplete region where $M_{\rm{UV}}$ is larger than the $M_{\rm{UV}}$ limit which corresponds to the $5\sigma$ NB limiting magnitude, 
assuming the typical $EW_0$ value of $69$\AA~which is the $e$-folding length of $EW_0$ distribution in the field (see \S3.3). For SDSS1426, we used the $J_{21}=10$ radius instead of the $J_{21}=1.0$ to evaluate the environmental effect because the radius of $J_{21}=1.0$ is larger than the FoV due to the brightness of the quasar. 
The upper limit for the case of no LAE detection 
\citep[i.e., corresponds to $N=1.841$; ][]{Gehrels86}
 is shown at $M_{\text{UV}}=-13.5$.   
 }\label{uv}
\end{figure*}

\begin{figure}
\begin{center} 
 \plotone{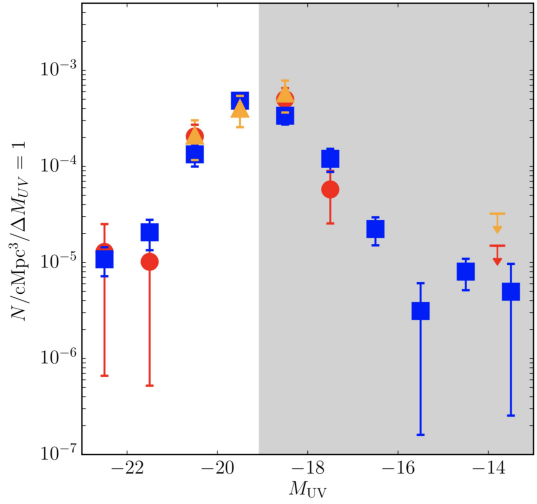}
\end{center}
\caption{The average distribution of UV luminosity of LAEs in the vicinity of quasars. 
The red circles and blue squares show the distribution of $M_{\rm{UV}}$ in the vicinity and outer region of quasars, respectively. 
The error bars show  the standard error of mean of the sample number in each bin. The gray shade region shows the completeness limit which corresponds to the $M_{\rm{UV}}$ at the shallowest $5\sigma$ NB limiting magnitude (that is, 24.95 in SDSS2131), where we assumed the typical $EW_0$ value of $69$\AA~which is the $e$-folding length of $EW_0$ distribution in the field (see \S3.3). The orange triangles show the distributions if we exclude the four most luminous quasars, SDSS1411, SDSS1426, OH91--121 and HB89--1835. 
The upper limit for no LAE detection 
 is shown at $M_{\text{UV}}=-13.8$. 
}\label{aveuv}
\end{figure}

The UV luminosity of a LAE is estimated using our LAE model (see Appendix A). 
Note that the derived $M_{\rm{UV}}$ distribution is affected by the discrepancy of completeness between NB and BB filters, since our Ly$\alpha$ selected sample is constructed from a NB magnitude limited sample. 
It is difficult to correct the sample in a completeness, and we only focus on the relative difference of the distributions found in the vicinity and outer regions of the quasars. 
Figure \ref{uv} shows the $M_{\rm{UV}}$ distribution of LAEs in the quasar fields. 
A deficit of faint ($M_{\rm{UV}}>-17$) LAEs in the proximity of a quasar can be seen in several fields. 
To see this trend more clearly, the average UV luminosity distribution for all quasar fields but SDSS1426 is shown in Figure \ref{aveuv}. 
We normalize the height of each distribution with that of SDSS1701 when taking the average 
because we want to see only the difference in the shapes of the distributions. 
The $M_{\rm{UV}}$ distributions are almost identical between the vicinity and outer region of quasars at the bright end $M_{\rm{UV}}<-17.0$, 
while at the end, faint LAEs are significantly deficient in the vicinity of the quasars. 
The result suggests that fainter LAEs with $M_{\rm{UV}}>-17.0$ tend to avoid the central quasar, as expected if the photoevaporation effects are shown.  
Moreover, 
we found that faint LAEs with $M_{\rm{UV}}>-18.0$ are only appeared around the four brightest quasars, SDSS1426, OH91--121, SDSS1411 and HB89--1835, 
whose bolometric luminosities are estimated to be brighter than that of a typical hyper luminous quasar, $L_{\text{bol}}>10^{47}$ erg/s \citep[][]{Bischetti17}, 
using the bolometric correction of \citet{Runnoe12}.  
If we tried to exclude these hyper luminous quasars, the trend becomes clearer as shown in Figure \ref{aveuv}.  
This trend did not change even if the two possible quasars in HB89--1835 and SDSS1250 (white stars in Figure \ref{colormag} and \ref{density} ) are included. 
However, there is an uncertainty that the difference can be seen only beyond the completeness limit (vertical gray line in Figure \ref{aveuv}).

\begin{figure*}
\begin{center} 
 \plotone{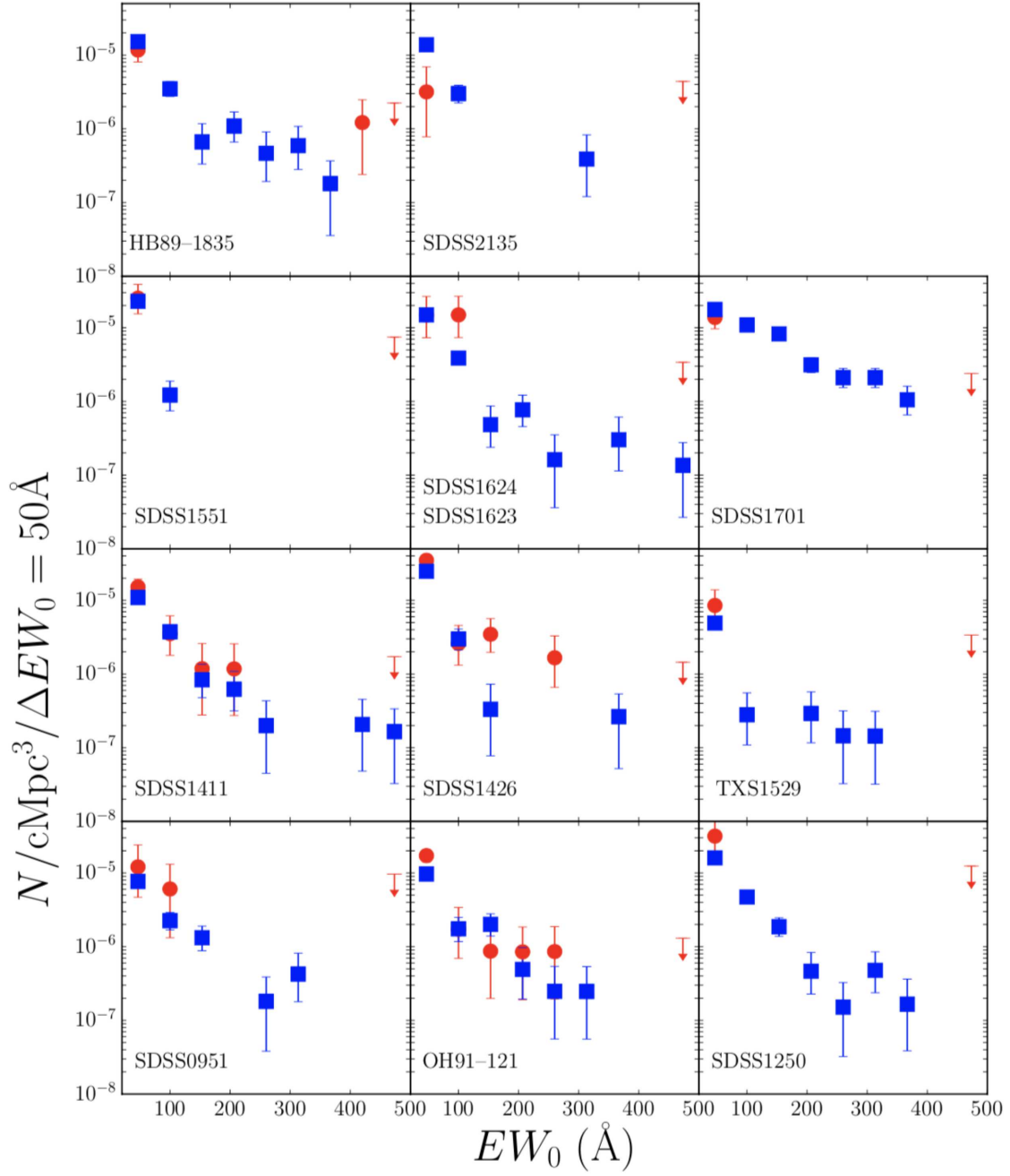}
\end{center} 
\caption{Distribution of $EW_0$ of LAEs in the vicinity of quasars. 
The filled red circles and blue squares show the distribution in the vicinity and outer region of quasars, respectively. 
The error bar assumes a binomial distribution \citep[][]{Gehrels86}. For SDSS1426, we used the $J_{21}=10$ radius instead of the $J_{21}=1.0$ to evaluate the environmental effect because the radius of $J_{21}=1.0$ is larger than the FoV due to the brightness of the quasar.  
The upper limit for the case of no LAE detection 
\citep[i.e., corresponds to $N=1.841$; ][]{Gehrels86}
 is shown at $EW_0=470$\AA.   
}\label{ew}
\end{figure*}

\begin{figure}
\begin{center} 
 \plotone{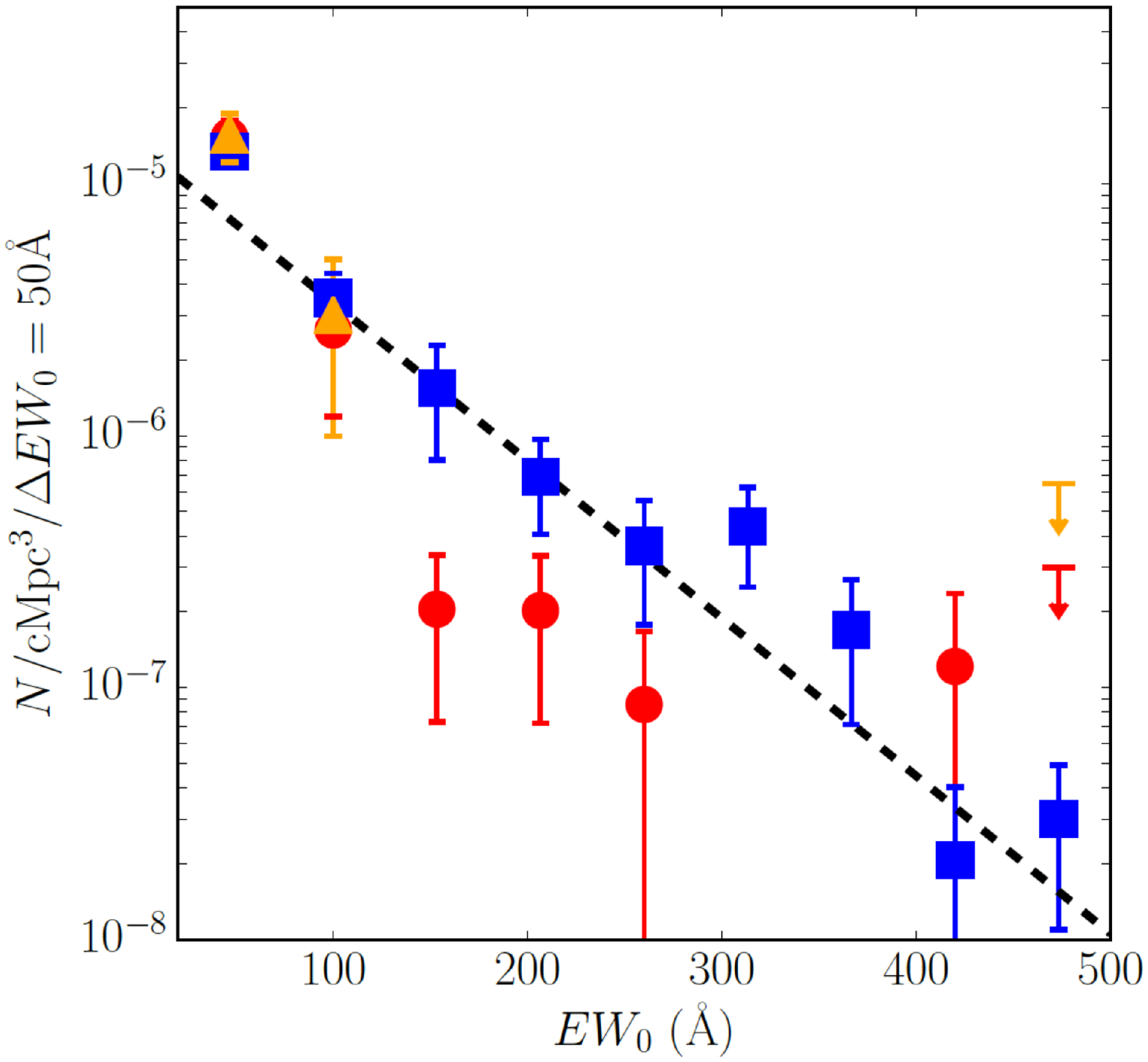}
\end{center}
\caption{ The average distribution of $EW_0$ of LAEs in the vicinity of quasars. 
The red circles and blue squares show the distribution in the vicinity and outer region of quasars, respectively. 
The error bars show the standard error of mean of the sample number in each bin. The dashed line indicates the best fit of the distribution in the outer region. The orange triangles show the distributions in the vicinity but for four of the most luminous quasars, SDSS1411, SDSS1426, OH91--121 and HB89--1835. 
The upper limit for no LAE detection is shown at $EW_0=470$\AA.   
}\label{aveew}
\end{figure}

\subsection{Equivalent Width Distributions of LAEs in the Vicinity of Quasars}
It is well-known that the $EW_0$ of Ly$\alpha$ emission is anti-correlated with the stellar masses of LAEs \citep[e.g.][]{Nilsson09}. 
The $EW_0$ of a LAE is estimated based on our LAE model (see Appendix A). 
The correlation coefficient $\rho$ and $P$-value  between the $EW_0$ and $M_{\rm{UV}}$ distribution are 0.74 and $\ll0.05$ 
in the Spearmann rank correlation test. 
The distribution of the $EW_0$ of Ly$\alpha$ emission of LAEs for each field is shown in Figure \ref{ew}. 
Figure \ref{ew}  suggests that the LAEs in the quasar proximity region tend to have lower $EW_0$, especially, in the case of SDSS1701. 
The average $EW_0$ distribution for all fields but SDSS1426 are shown in Figure \ref{aveew} in order to make the trend more clear. 
We found a tendency that LAEs with $EW_0>150$\AA~are scarcer in the proximity of quasars, 
 while the abundance of LAEs with $EW_0\lesssim 100$\AA~is almost the same in the vicinity and outer region.  
In the quasar proximity region, high-$EW_0$ LAEs with $EW_0>150$\AA~are appeared only around the four most luminous quasars.  
As is the case with the $M_{UV}$ distribution, 
the trend becomes clearer if these hyper luminous quasars are excluded (Figure \ref{aveew}). 
The $EW_0$--UV luminosity relation \citep[][]{Nilsson09} suggests that a LAE with high $EW_0$ tends to have lower UV luminosity. 
In fact, the $M_{\rm{UV}}$ of $>-18.0$ corresponds to $EW_0>150$\AA~in our $EW_0$--UV luminosity relation. 
Our finding of a paucity of LAEs with high $EW_0$, which corresponds to low stellar mass, objects around quasars could be caused by the quasar photoevaporation effect. 
More quantitative discussion is made in \S 4.1.  
Again, even if we add the two possible quasars (white stars in Figure \ref{colormag} and \ref{density} ),  the result did not change. 

The deficit of LAEs with high $EW_0$ in the vicinities of our quasars could be just 
due to the limited area rather than photoevaporation effects.  
We define $N^{\rm{obs}}_{<r_{\rm{prox}}}(>\!150$\AA$)$ and $N^{\rm{exp}}_{<r_{\rm{prox}}}(>\!150$\AA), which are summarized in Table \ref{t3}, as the  observed and expected completeness-corrected number of LAEs with $EW_0>150$\AA, respectively. 
$N^{\rm{exp}}_{<r_{\rm{prox}}}(>\!150$\AA) is estimated assuming that the intrinsic LAE number density in the vicinity of a quasar is the same as that in the outer region.  
The highest expected number $N^{\rm{exp}}_{<r_{\rm{prox}}}(>\!150$\AA)$=12.8$ was found in the SDSS1701 field even though $N^{\rm{obs}}_{<r_{\rm{prox}}}(>\!150$\AA)$ = 0$. 
The difference may occur due to two possible overdense regions not centered on the SDSS1701 quasar. 
As seen in Table \ref{t3}, the total $N^{\rm{obs}}_{<r_{\rm{prox}}}(>\!150$\AA$)$ is significant smaller than the total $N^{\rm{exp}}_{<r_{\rm{prox}}}(>\!150$\AA$)$. 
Interestingly, there are no LAEs with $EW_0>150$\AA~ in the vicinity except for the four brightest quasars. 
This is unlikely to be caused by a possible optical vignetting of S-Cam because our quasars always lie in the center of the images. 
We also perform $\chi^2$ fitting to the equation, $e^{-EW_0/w_0}$ that describes the $EW_0$ distribution in the outer region, 
and got $w_0 = 69^{+4}_{-6}$\AA, which is consistent with \citet{Guaita10} and \citet{Gronwall07} 
who found that $w_0 = 83^{+10}_{-10}$\AA~at $z\!\sim\!2$ and $w_0 = 76^{+11}_{-8}$\AA~at $z\!\sim\!3$, respectively. 
On the other hand, the decay scale $w_0$ estimated by \citet{Nilsson09a} and \citet{Mawatari12} at $z\!\sim\!2$ is 
 lower, $w_0 = 48.5^{+1.7}_{-1.7}$\AA~and $43.7^{+0.43}_{-0.43}$\AA, respectively, and they suggested more redder galaxies populate LAEs at lower redshift. 
If we focus on $z\!\sim\!2$ in our sample, we get $w_0 = 73^{+4}_{-6}$\AA. 
While some in $w_0$ evolution is observed at a redshift range of $\sim3-6$ \citep[][]{Hashimoto17b}, 
a consistent picture still needs to be derived for $z\sim2-3$.

\section{Discussion}
\citet{Kashikawa07}  showed that a gas cloud in a halo with a small dynamical mass $M_{\rm{vir}}\lesssim 10^9 M_\odot$ will experience a considerable delay in star formation under a local UV background with $J_{21}\gtrsim1$ based on the hydrodynamical simulation of \citet{Kitayama00,Kitayama01}.  
Recently, \citet{Hashimoto17} concluded that 
LAEs with large Ly$\alpha$, $EW_0\sim200-400$\AA, have 
notably small stellar masses of $10^{7-8} M_\odot$, the median value of which is $7.1^{+4.8}_{-2.8} \times 10^7 M_\odot$ using the SED fitting method.     
Also, the stellar masses of LAEs with a smaller $EW_0$ of $\lesssim150$\AA, are found to be $5.9^{+19.2}_{-2.2}\times10^{8} M_\odot$ \citep[][]{Shimakawa17}.  
The stellar mass of LAEs can also be estimated from the Star Formation Rate (SFR)-stellar mass relation, the ``main sequence". 
The average SFRs is $0.9$ $M_{\odot}$ yr$^{-1}$ for LAEs with $EW_0>150$\AA~and $6.4$ $M_{\odot}$ yr$^{-1}$ for those with $EW_0<150$\AA \citep[][]{Kennicut98}.  
Based on the main sequence of LAEs by \citet{Vargas14}, the stellar mass is estimated as $10^{7-8} M_\odot$ and $10^{8-9} M_\odot$ for $EW_0>150$\AA~and $EW_0<150$\AA, respectively, which is consistent with the SED fitting result. 
The stellar to halo mass ratio (SHMR) of LAEs at $z\!\sim\!2$ is estimated to be $0.02_{-0.01}^{+0.07}$, assuming that each halo hosts one galaxy \citep[][]{Kusakabe17}. 
Therefore, it is expected that LAEs with $EW_0\gtrsim150$\AA~have a halo mass $M_{\rm{h}}=3.6^{+12.7}_{-2.3}\times10^9 M_{\odot}$, and LAEs with $EW_0\lesssim150$\AA~occupy more massive halos of $M_{\rm{h}}=2.9^{+14.0}_{-1.8}\times10^{10} M_\odot$, using the stellar mass estimated by \citet{Hashimoto17} and \citet{Shimakawa17}, and the SHMR of \citet{Kusakabe17}. 
Note that the typical age of LAEs with $EW_0>150$\AA~is $\lesssim20$ Myr \citep[][]{Hashimoto17}, which is shorter than  
the predicted delay time of $>20$ Myr in star formation under a local UV background with $J_{21} \gtrsim 1$ for galaxies with the halo mass of $3.6\times10^9 M_{\odot}$ also based on hydrodynamical simulations \citep[][]{Kashikawa07}. 
Photoionization heating, which can raise the gas temperature, by the strong UV radiation from a quasar delays star formation in a small halo with mass smaller than $\sim 3\times10^9 M_{\odot}$ and its typical delay time is expected to be $>20$ Myr. 
In other words, young galaxies with an age of $<20$ Myr are prevented from forming by the quasar photoionization feedback. 
Therefore, the quasar photoionization feedback can reasonably explain our finding of a deficiency of LAEs, especially for those LAEs having high $EW_0$ and low mass halos in the proximity of quasars. 
It should be noted that quasar life time is also related to the effect.  
Even low-mass galaxies could collapse if they form before the quasar active phase. 
Also, if the  quasar life time is longer than $\sim20$ Myr, a possible star formation delay mentioned above will be washed out.  
Interestingly, the delay time is comparable to the  a fiducial value of quasar lifetime of $10^{7.5}$ yr, which is in agreement with observational results 
\citep[][]{Martini04, Shen07} and models \citep[][]{Hopkins06}, though with a large uncertainty.

\citet{Uchiyama2017}  found that the number density of $g$-dropout galaxies around SDSS quasars, whose median UV luminosity, log $L_{912}$, is equivalent to this study, tends to be slightly deficient at $<0.5$ pMpc around quasars. 
This might also be caused by the quasar photoionization feedback, though the scale is smaller than the proximity size, $r_{\rm{prox}} = 1.64_{-3.0}^{+3.0}$ pMpc, probed by LAEs in the study.  


On the other hand, we were not able to find evidence of the photoevapolation effect in four hyper quasar fields, where the effect should perhaps be the strongest. 
According to the quasar evolution model of \citet{Kawakatu09}, the brightness of a quasar is expected to monotonically decrease during the active phase.   
Note that  SDSS1426 has a large total infrared luminosity, $L_{\rm{IR}}=10^{14.29}$ \citep[][]{Weedman12}, meaning that it is a X-ray bright, optically normal galaxies (XBONG) \citep[][]{Schawinski15} or dusty quasar which is thought to be in the early stage of the quasar phase in the merger scenario  \citep[][]{Kauffmann2002}.  
The ionization feedback from the very luminous quasars might have not yet affected galaxies in their vicinity because they just appeared.   
The result suggests that the timescale which quasars can affect the gas cooling is estimated to be at most $\sim20$ Myr, 
which is shorter than the quasar active phase time scale of a few $\times 100$ Myr \citep[][]{Kawakatu09}, because the typical age of a LAE with high $EW_0$ is $\lesssim20$ Myr \citep[][]{Hashimoto17}. 
In addition, radio-loud quasars were found to reside in almost the same environment as radio-quiet quasars, as observed by \citet{Donoso10} at $z\sim0.5$. 
We further found that quasar photoionization feedback is independent of radio-loudness.  

In the context of  the AGN unification model, quasars are observed relatively face-on with respect to the AGN torus, suggesting that the quasar photoionization feedback cannot affect directions that are transverse to the line of sight. 
The type 1 fraction of the AGN population is estimated to be around $50-70$\% \citep[][]{Simpson05}, 
by estimating the O[III] luminosity from UV luminosity by the bolometric correction factors of \citet{Runnoe12} and  \citet{Shen11}. 
Assuming that the fraction corresponds to the quasar viewing angle, the radiation solid angle of quasars is $0.47$ -- $0.79$ radian. 
Thus, the expected number of LAEs  with $EW_0>150$\AA, which happen to lie in the region where the quasar UV radiation is obscured by the torus, is $7.3$ -- $12.1$ based on the average $EW_0$ distribution (Figure \ref{aveew}). 
This expectation number is consistent with the observed number of LAEs with $EW_0>150$\AA~in the quasar proximity regions ($N^{\rm{obs}}_{<r_{\rm{prox}}}(>\!150$\AA) $= 7.15$). 
In other words, the observed number of LAE with $EW_0>150$\AA~in the quasar proximity regions can be explained by the possible anisotropic radiation field of quasars. 


We compared our results with \citet{Marino17} who investigated the photoionization effect in six quasar fields targeting galaxies with strong Ly$\alpha$ emission at $3<z<4$ using MUSE. 
They found an opposite, positive correlation suggesting that LAEs with high $EW_0$ tend to cluster near the quasars. 
However, most of these high Ly$\alpha$ EW objects are likely to be caused by faint fluorescence, which can be detected by MUSE with high sensitivity. 
If their sample was confined to objects with UV continuum detections, 
there is no significant difference in the $EW_0$ distributions in the vicinity and outer regions of their quasars.  
Five of the six quasars are very luminous quasars, $M_B\sim-30.0$ \citep[][]{Veron10}, similar to SDSS1426 in our sample, 
suggesting that our result is consistent with their results. Even if we limit our SDSS1426 LAEs to the sample with UV continuum detections only, this 
trend does not change.    
We found 195 LAEs, which were not detected in BB images. 
The fraction of these continuum-undetected LAEs to the total 1171  sample is small (17\%). 
Even if the 195 LAEs were excluded, the $EW_0$ and UV luminosity distribution did not change within 1$\sigma$  errors. 
There is no difference in the number density of the continuum-undetected LAEs between the quasar vicinity and the outer region. 
Their continuum flux might not detected due to the shallowness of BB images, while they could be caused by the quasar fluorescence, which will be discussed in a forthcoming paper.

Another possible scenario to explain the scarcity of high $EW_0$ LAE, which is expected to be much younger than those with lower $EW_0$ 
\citep[][]{Hashimoto17}, is that quasar fields could predominantly contain a more evolved population, such as LBGs. 
If quasars favor overdense environments, which enhance early galaxy formation, all the LAEs around quasars might have already evolved into LBGs. 
This is also related to the quasar duty cycle:  LBGs might have already formed at the time of the active phase of the quasar host galaxy.
However,  our recent statistical study showed that the most luminous quasars tend to avoid the overdense regions of LBGs, 
suggesting that quasars are not hosted by very massive halos that lack high $EW_0$ LAEs alone \citep[][]{Uchiyama2017}. 
This result raises some doubts on the above scenario. 
Simultaneous sampling for both LAEs and LBGs for quasar fields are required to validate this hypothesis.

\section{Conclusion}
We have carried out deep and wide imaging targeting 11 quasar fields to systematically study the photoevaporation effect for young and low mass galaxies, LAEs. 
In order to examine variation of the photoionization effects, 
we selected quasars with a range of properties, such as radio-loudness, black hole mass (log$ M_{\rm{BH}} /M_{\odot}=8.59-9.89$), and luminoisty (log$\lambda L_\lambda (912$\AA )$=45.6-47.8$).   
We selected LAEs in the quasar fields, and obtained 1171 LAEs in total up to 5$\sigma$ NB limiting magnitudes ($\sim25-27$) and carefully checked for fake detections by eye.  
The proximity zone of a quasar is defined by the region where the local UV radiation from the quasar is comparable to the UV background. 

We obtained the following results. 
\begin{itemize}
\item The 11 quasars tend to reside in average LAE density environments, whose average overdensity $\delta$ is $0.48$ with $0.82\sigma$ significance. 
One quasar pair in our sample appears in the most overdense region. These findings are consistent with previous findings based on LBGs at $z\!\sim\!4$ \citep[][]{Onoue17a, Uchiyama2017}.   
\item We compared the $EW_0$ and $M_{\rm{UV}}$ distribution of LAEs in the vicinity and outer region of a quasar. 
We found that LAEs with high $EW_0$ ($EW_0\gtrsim150$\AA~) or equivalently, faint UV luminosity ($M_{\rm{UV}}\gtrsim-17.0$), are relatively scarce in the vicinity of the quasars. 
The range of $EW_0$ or $M_{\rm{UV}}$ corresponds to notably smaller halo mass of $M_{\text h}\sim 10^{9-10} M_\odot$ estimated either from  SED fitting or  the main sequence, assuming $M_{\text h}/M_*\sim50$. 
The LAEs with such low halo mass are expected to be subjected to quasar photoevaporation.    
\item Counter to the main trend, we find that the feedback seems to be less effective in fields with hyper luminous quasars, 
but this could be explained if these luminous quasars are in too early stage of quasar activity to affect the gas cooling.  
The environment around radio loud quasars is similar to that around radio quiet quasars.  
\end{itemize} 

We, for the first time, performed a systematic study of quasar negative feedback on the surrounding galaxies. 
Future cosmological barionic simulations including this kind of feedback 
will be invaluable for the interpretation of our results.  
Finally, we note that another important feedback process, namely quasar fluorescence, will be discussed in a forthcoming paper.  
 
\acknowledgments 
We are grateful to Masao Hayashi, Takuma Izumi and Taiki Kawamuro for useful comments and discussion.  
We thank the referee for his/her helpful comments that improved the manuscript.  
This work was partially supported by Overseas Travel Fund for Students (2016) of the Department of Astronomical Science, SOKENDAI (the Graduate University for Advanced Studies). 
NK acknowledges support from the JSPS grant 15H03645. RAO received support from CNPq and the Visiting Scholar Program of the Research Coordination Committee of the National Astronomical Observatory of Japan (NAOJ). 

\appendix
\section{Our model for Lyman $\alpha$ emitters}
Here, we present our model of a LAE used to estimate the continuum and Ly$\alpha$ emission luminosity from the galaxy based on BB and NB imaging. 
The continuum flux density $f^{\rm{cont}}_{\nu}$ is assumed as 
\begin{eqnarray}
f^{\rm{cont}}_{\nu} &=& a (\nu - \nu_{\rm{Ly\alpha}}) + b ~~~\rm{at}~~~ \nu < \nu_{\rm{Ly\alpha}}, \\
                       &=& 0   ~~~~~~~~~~~~~~~~~~~~~~\rm{at}~~~ \nu \ge \nu_{\rm{Ly\alpha}}, 
\end{eqnarray}
where  $\nu_{\rm{Ly\alpha}}$ indicates the observed-frame frequency at the Ly$\alpha$ emission, and $a$ and $b$ are constants. 
We discuss two different cases : (1) BB covers Ly$\alpha$ emission, and (2) BB does not cover the emission. 

\subsubsection{Case 1 :  BB covers Ly$\alpha$ emission}
The observed flux density in the NB and BB are, to a first approximation, given by 
\begin{eqnarray}
\left< f_{\nu} \right>_{\rm{NB}} \Delta_{\rm{NB}}&=& F_{\rm{Ly\alpha}} + \frac{b \Delta_{\rm{NB}}}{2}  \label{A1}\\
\left< f_{\nu} \right>_{\rm{BB}} \Delta_{\rm{BB}}&=& F_{\rm{Ly\alpha}} + b \delta + \frac{-a \delta^2}{2} \label{A2}\\
\delta &\equiv& \nu_{\rm{Ly\alpha}} - \nu_{\rm{BB}} + \frac{\Delta_{\rm{BB}}}{2}, 
\end{eqnarray}
respectively. Here, $F_{\rm{Ly\alpha}}$ is the flux of Lyman $\alpha$ emission, and  $\Delta_{\rm{NB}}$ 
and $\Delta_{\rm{BB}}$ are the FWHM of the NB and BB filter in frequency space, respectively, and 
$\nu_{\rm{BB}}$ is the central frequency of the BB filter. In addition, we impose the following boundary condition, 
\begin{eqnarray}
m_{\rm{BB}}-m_{\rm{NB}} = h ~~~\rm{at}~~~F_{\rm{Ly\alpha}} = 0. \label{A3}
\end{eqnarray}
The $h$ indicates the color term (see \S2). 

Solving the equation (\ref{A1}), (\ref{A2}) and (\ref{A3}), we can get a Ly$\alpha$ luminosity $L_{\rm{Ly\alpha}}$, a UV flux density $f_{\rm{UV}}$,  
a rest-frame $EW_0$ of Ly$\alpha$ emission $EW_0$, and $f(EW_0 =20$\AA$)$ in equation (\ref{eq1}) as follows;

\begin{eqnarray}
&L_{\rm{Ly\alpha}}       & =  \frac{4 \pi d_l^2 \Delta_{\rm{BB}} \Delta_{\rm{NB}}}{\Delta_{\rm{NB}} - 10^{-0.4 h} \Delta_{\rm{BB}}} \left( \left< f_{\nu} \right>_{\rm{BB}} - 10^{-0.4 h}  \left< f_{\nu} \right>_{\rm{NB}}\right)       \\
&f_{\rm{UV}} &= 2 \frac{\Delta_{\rm{NB}}  \left< f_{\nu} \right>_{\rm{NB}} - \Delta_{\rm{BB}}  \left< f_{\nu} \right>_{\rm{BB}} }{\Delta_{\rm{NB}} - \Delta_{\rm{BB}} 10^{-0.4 h}} \\
&EW_0                  &=  \frac{(1+z)\lambda_{\rm{Ly\alpha}}^2}{2c} \frac{\Delta_{\rm{NB}}\Delta_{\rm{BB}}  \left( \left< f_{\nu} \right>_{\rm{BB}} - 10^{-0.4 h}  \left< f_{\nu} \right>_{\rm{NB}}\right)}{\left< f_{\nu} \right>_{\rm{NB}} \Delta_{\rm{NB}} - \left< f_{\nu} \right>_{\rm{BB}} \Delta_{\rm{BB}}}    \\
&f                           &= -2.5 \log_{10} \left[ \frac{2c EW_0 \Delta_{\rm{NB}} + 10^{-0.4h} (1+z) \lambda_{\rm{Ly\alpha}}^2  \Delta_{\rm{NB}} \Delta_{\rm{BB}}}{2c EW_0 \Delta_{\rm{BB}} + (1+z) \lambda_{\rm{Ly\alpha}}^2  \Delta_{\rm{NB}} \Delta_{\rm{BB}}} \right]
\end{eqnarray}
, where $z$ is the redshift of the LAE, $\lambda_{\rm{Ly\alpha}}$ is the rest-frame wavelength of Ly$\alpha$ emission, and $c$ is speed of light. 
$d_l$ is the luminosity distance at each redshift. 

\subsubsection{Case 2 :  BB does not cover Ly$\alpha$ emission}
The observed flux density in the NB and BB are obtained by 
\begin{eqnarray}
\left< f_{\nu} \right>_{\rm{NB}} \Delta_{\rm{NB}}&=& F_{\rm{Ly\alpha}} + \frac{b \Delta_{\rm{NB}}}{2}  \label{A4}\\
\left< f_{\nu} \right>_{\rm{BB}} \Delta_{\rm{BB}}&=&  \left(a(\nu_{\rm{Ly\alpha}} - \nu_{\rm{BB}}) +b \right) \Delta_{\rm{BB}}   \label{A5} \\
m_{\rm{BB}}-m_{\rm{NB}} &=& h ~~~\rm{at}~~~F_{\rm{Ly\alpha}} = 0 \label{A6} 
\end{eqnarray}
as in Case 1. 
Then, we get the following equations : 
\begin{eqnarray}
&L_{\rm{Ly\alpha}}&        =  4 \pi d_l^2 \Delta_{\rm{NB}} \left( 10^{0.4 h} \left< f_{\nu} \right>_{\rm{BB}} -  \left< f_{\nu} \right>_{\rm{NB}} \right)       \\
&f_{\rm{UV}}& = 2  \left< f_{\nu} \right>_{\rm{BB}} 10^{0.4 h} \\
&EW_0           &       =  \frac{(1+z)\lambda_{\rm{Ly\alpha}}^2}{2c} \frac{\left( 10^{-0.4 h}  \left< f_{\nu} \right>_{\rm{NB}} - \left< f_{\nu} \right>_{\rm{BB}} \right)\Delta_{\rm{NB}}}{\left< f_{\nu} \right>_{\rm{BB}} }    \\
&f                    &       = -2.5 \log_{10} \left[\frac{(1+z) \lambda_{\rm{Ly\alpha}}^2 \Delta_{\rm{NB}} 10^{-0.4 h} }{2c EW_0 + (1+z)\lambda_{\rm{Ly\alpha}}^2 \Delta_{\rm{NB}} } \right]. 
\end{eqnarray}

We have assumed that the flux density of a LAE is zero at wavelengths shorter than the Ly$\alpha$ wavelength. 
Even though, the residual flux could be as high as $\sim60$\% \citep{Madau95}.  
In that case, the $EW_0$ estimate increases by 60\%.  

\end{document}